\documentclass[reprint,showpacs,preprintnumbers,amsmath,amssymb,aps,prb,floatfix]{revtex4-1}
\usepackage[usenames]{color}
\usepackage{graphicx}
\usepackage{dcolumn}
\usepackage{bm}

\begin{document}

\title{Spin-flipping with Holmium: Case study of proximity effect in
  superconductor/ferromagnet/superconductor heterostructures}

\author{Daniel Fritsch} 

\author{James F. Annett} 

\affiliation{H. H. Wills Physics Laboratory, School of Physics,
  University of Bristol, Bristol BS8 1TL, UK}

\date{\today}

\begin{abstract}
Superconductor/ferromagnet/superconductor heterostructures exhibit a
so-called long-range proximity effect provided some layers of conical
magnet Holmium are included in the respective interface regions. The
Ho layers lead to a spin-flip process at the interface generating
equal-spin spin-triplet pairing correlations in the ferromagnet. These
equal-spin spin-triplet pairing correlations penetrate much further
into the heterostructure compared to the spin-singlet and unequal-spin
spin-triplet correlations which occur in the absence of Ho. Here we
present calculations of this effect based on the spin-dependent
microscopic Bogoliubov$-$de Gennes equations solved within a
tight-binding model in the clean limit. The influence of the
ferromagnet and conical magnet layer thickness on the induced
equal-spin spin-triplet pairing correlations is obtained and compared
to available experimental data. It is shown that, in agreement with
experiment, a critical minimum thickness of conical magnet layers has
to be present in order to observe a sizeable amount of equal-spin
spin-triplet pairing correlations.
\end{abstract}

\pacs{Valid PACS appear here} 

\maketitle

\section{Introduction}
\label{Introduction}

The proximity effect arises because superconducting pairing
correlations are able to penetrate the interface between a
superconductor (SC) and a nonmagnetic metal. The correlations decay in
the adjacent metal with a characteristic length scale. Replacing the
nonmagnetic metal by a ferromagnet (FM) the pairing correlations
become oscillatory and this characteristic length scale is also
considerably reduced. The penetration depth of the spin-singlet
pairing correlation into the ferromagnetic layer depends on the
exchange interaction, which also leads to FFLO-like
oscillations~\cite{Fulde_PhysRev135_A550,Larkin_JETP20_762} visible in
the ferromagnetic region. The SC/FM interface also generates
unequal-spin spin-triplet pairing correlations which have the same
oscillating and decaying pattern as for the spin-singlet
correlations. However, according to a theoretical prediction by
Bergeret \textit{et al.}~\cite{Bergeret_PRL86_4096} it should be
possible to also create equal-spin spin-triplet pairing correlations,
provided the interface allows for some kind of spin-flip process. In
principle, these equal-spin spin-triplet pairing correlations should
be compatible with a ferromagnetic exchange interaction and allow for
much larger penetration depth in the ferromagnetic region of the
heterostructure compared to the spin-singlet proximity effect. This
so-called long-range proximity effect has stimulated a lot of
experimental and theoretical work, and has been summarised in reviews
by Buzdin~\cite{Buzdin_RMP77_935} and Bergeret \textit{et
  al.}~\cite{Bergeret_RMP77_1321}

From the experimental point of view several possible sources of
spin-flip processes have been identified and have been realised in
heterostructure setups. It should be noted that experimental evidence
for the long-range proximity effect mostly stems from observations of
super currents in SC/FM/SC Josephson junctions with FM length scale
incompatible with the spin-singlet proximity effect. Experimentally
the heterostructures which have been already realised include
half-metallic
metals,~\cite{Keizer_Nature439_825,Anwar_PRB82_100501,Visani_NatPhys8_539}
introducing a magnetic inhomogeneity at the
interface,~\cite{Khaire_PRL104_137002} noncollinear magnetic
interfaces,~\cite{Klose_PRL108_127002,Gingrich_PRB86_224506,Zdravkov_PRB87_144507}
and helical~\cite{Halasz_PRB84_024517} or conical magnetic
structures,~\cite{Sosnin_PRL96_157002,Halasz_PRB79_224505,Robinson_Science329_59}
to name but a few.

On the theoretical side, several basically very different approaches
have been applied to particular heterostructures to investigate
equal-spin spin-triplet pairing correlations and how they are affected
by specific spin-flip mediating interfaces. In line with the
experimental observations various interfaces which have been studied
include half-metallic
metals,~\cite{Eschrig_PRL90_137003,Eschrig_NaturePhys4_138}
inhomogeneous
magnetisations,~\cite{Bergeret_PRL86_4096,Alidoust_PRB81_014512}
noncollinear magnetisations,~\cite{Volkov_PRL90_117006} and helical
(or conical)
magnets.~\cite{Halasz_PRB79_224505,Wu_PRB86_184517,Fritsch_NJP16_055005,Fritsch_JPCM26_274212}
Green's function techniques based on solutions of the
Eilenberger~\cite{Eschrig_PRL90_137003} and the Usadel equation have
been
reported,~\cite{Bergeret_PRL86_4096,Volkov_PRL90_117006,Fominov_PRB75_104509,Alidoust_PRB81_014512,Kawabata_JPSJ82_124702}
as well as self-consistent solutions of the Bogoliubov$-$de Gennes
(BdG) equations for suitable tight-binding
models.~\cite{Halterman_PRL99_127002,Halterman_PRB77_174511,Wu_PRB86_184517,Fritsch_NJP16_055005,Fritsch_JPCM26_274212}

Here we present results based on a model heterostructure consisting of
a $s$-wave SC/FM/SC junction with additional conical magnet (CM)
layers introduced at the interfaces. The relevant tight-binding model
is solved in the microscopic spin-dependent BdG equations and
solutions are iterated to self-consistency. From the respective
eigenfunctions we obtain the different spin-triplet pairing
correlations and discuss the influence of CM and FM layer
thickness. We find that, in agreement with experimental observations,
a critical minimal thickness of CM layers has to be present to observe
the long-range proximity effect.

The paper is organised as follows. Sec.~\ref{Sec2} provides
theoretical background, including the BdG equations and
heterostructure setup in Sec.~\ref{Sec2_1}, and the spin-triplet
pairing correlations in Sec.~\ref{Sec2_2},
respectively. Sec.~\ref{Sec3} is devoted to results, where the
influence of FM (CM) thickness on spin-triplet pairing correlations is
discussed in detail in Sec.~\ref{Sec3_1} (Sec.~\ref{Sec3_2}). A
summary and concluding remarks are given in
Sec.~\ref{SummaryAndOutlook}.

\begin{widetext}

\section{Theoretical background}
\label{Sec2}

\subsection{Method, heterostructure setup and computational details}
\label{Sec2_1}

The results presented here are based on self-consistent solutions of
the microscopic BdG equations in the clean limit. For the
spin-dependent case and incorporating the vector components of a
general exchange field ${\bf h}$ the BdG equations
read~\cite{Fritsch_NJP16_055005,Annett_Book,KettersonSong}
\begin{eqnarray}
  \label{EqBdGGeneral} \left(
  \begin{array}{cccc}
    {\cal H}_{0} - h_{z} & -h_{x} + i h_{y} & \Delta_{\uparrow
      \uparrow} & \Delta_{\uparrow \downarrow} \\ - h_{x} - i h_{y}
    & {\cal H}_{0} + h_{z} & \Delta_{\downarrow \uparrow} &
    \Delta_{\downarrow \downarrow} \\ \Delta_{\uparrow \uparrow}^{*}
    & \Delta_{\downarrow \uparrow}^{*} & -{\cal H}_{0} + h_{z} &
    h_{x} + i h_{y} \\ \Delta_{\uparrow \downarrow}^{*} &
    \Delta_{\downarrow \downarrow}^{*} & h_{x} - i h_{y} & -{\cal
      H}_{0} - h_{z}
  \end{array}
  \right) \left(
  \begin{array}{c}
    u_{n\uparrow} \\ u_{n\downarrow} \\ v_{n\uparrow}
    \\ v_{n\downarrow}
  \end{array}
  \right) =\varepsilon_{n} \left(
  \begin{array}{c}
    u_{n\uparrow} \\ u_{n\downarrow} \\ v_{n\uparrow}
    \\ v_{n\downarrow}
  \end{array}
  \right),
\end{eqnarray}
\end{widetext}
with $\varepsilon_{n}$, and $u_{n\sigma}$ and $v_{n\sigma}$ denoting
the eigenvalues, and quasiparticle and quasihole amplitudes for spin
$\sigma$, respectively. The tight-binding Hamiltonian ${\cal H}_{0}$
can be simplified according
to~\cite{Fritsch_NJP16_055005,Covaci_PRB73_014503}
\begin{equation}
  \label{EqTBHamiltonianLinear}
        {\cal H}_{0} = -t \sum_{n}{ \left( c_{n}^{\dagger}c_{n+1} +
          c_{n+1}^{\dagger}c_{n} \right) } + \sum_{n}{ \left(
          \varepsilon_{n} - \mu \right) c_{n}^{\dagger} c_{n} }.
\end{equation}
At multilayer index $n$ we have the electronic creation
($c_{n}^{\dagger}$) and destruction operators ($c_{n}$), whereas the
next-nearest neighbour hopping parameter and the chemical potential
(Fermi energy) are chosen to be $t = 1$ and $\mu = 0$, respectively.

According to Balian and
Werthamer~\cite{Balian_PhysRev131_1553,Sigrist_RMP63_239} the general
form of the pairing matrix in Eq.~(\ref{EqBdGGeneral}) can be
rewritten as
\begin{equation}
  \label{EqPairingMatrix}
  \left( \begin{array}{cc} \Delta_{\uparrow \uparrow} &
    \Delta_{\uparrow \downarrow} \\ \Delta_{\downarrow \uparrow} &
    \Delta_{\downarrow \downarrow} \end{array} \right) = \left( \Delta
  + {\hat {\bf \sigma}}{\bf d}\right) i {\hat \sigma}_{2} =
  \left( \begin{array}{cc} -d_{x} + i d_{y} & \Delta + d_{z} \\ -
    \Delta + d_{z} & d_{x} + i d_{y} \end{array} \right),
\end{equation}
with $\hat{\cdots}$ indicating a $2 \times 2$ matrix,
respectively. Making use of the Pauli matrices ${\hat {\bf \sigma}}$,
the superconducting order parameter is described by a singlet (scalar)
part $\Delta$ and a triplet (vector) part ${\bf d}$,
respectively. Defining ${\hat \Delta}$ to contain only the triplet
${\bf d}$-vector components yields
\begin{equation}
  \label{EqTripletPairingMatrix}
        {\hat \Delta} = \left( {\hat {\bf \sigma}}{\bf d}\right) i
        {\hat \sigma}_{2} = \left( \begin{array}{cc} -d_{x} + i d_{y}
          & d_{z} \\ d_{z} & d_{x} + i d_{y} \end{array} \right).
\end{equation}
Considering this general form of this triplet pairing matrix the
product ${\hat \Delta}{\hat \Delta}^{\dagger}$ can be written as
\begin{equation}
  \label{EqDDdagger}
        {\hat \Delta}{\hat \Delta}^{\dagger} = |{\bf d}|^{2} {\hat
          \sigma}_{0} + i \left( {\bf d}\times{\bf d}^{*} \right)
        {\hat {\bf \sigma}},
\end{equation}
with $|{\bf d}|$ describing the gap function and ${\bf d}\times{\bf
  d}^{*}$ being a measure for the Cooper pair spin magnetic moment,
respectively.

\begin{figure}[b]
  \centering
  \includegraphics[width=0.475\textwidth,clip]{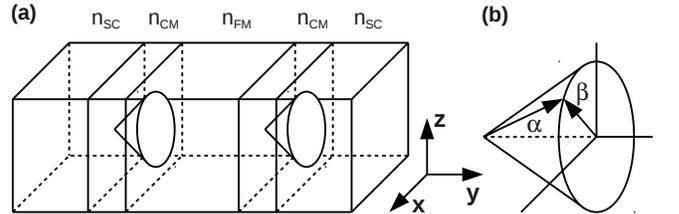}
  \caption{\label{Fig1} (a) Heterostructure setup (from left to
    right): spin-singlet $s$-wave superconductor ($n_{\rm SC}$
    layers), conical magnet ($n_{\rm CM}$ layers), ferromagnetic metal
    ($n_{\rm FM}$ layers), conical magnet ($n_{\rm CM}$ layers), and
    spin-singlet $s$-wave superconductor ($n_{\rm SC}$ layers). (b)
    Opening and turning angles $\alpha$ and $\beta$ of the conical
    magnet. According to Eq.~(\ref{EqHelicalMagnet}) $\alpha$ is
    measured from $+y$ towards $+z$, whereas $\beta$ is measured from
    $+z$ towards $+x$.}
\end{figure}

However, here we are interested in the case of $s$-wave SC
only. Therefore the pairing potential is restricted to a scalar
quantity $\Delta$ fulfilling the self-consistency condition
\begin{equation}
  \begin{split}
    \Delta({\bf r}) = & \frac{g({\bf r})}{2} \sum_{n}{ \bigl(
      u_{n\uparrow}({\bf r})v_{n\downarrow}^{*}({\bf r})
      [1-f(\varepsilon_{n})] \bigr. } \\ & + u_{n\downarrow}({\bf
      r})v_{n\uparrow}^{*}({\bf r}) f(\varepsilon_{n}) \bigl. \bigr),
  \end{split}
\end{equation}
where we are summing only over positive eigenvalues $\varepsilon_{n}$
and $f(\varepsilon_{n})$ denotes the Fermi distribution function
evaluated as a step function for zero temperature. Setting up the
multilayer structure as shown in Fig.~\ref{Fig1}(a) the effective
superconducting coupling parameter $g({\bf r})$ equals $1$ in the
$n_{\rm SC} = 250$ layers of spin-singlet $s$-wave superconductor to
the left and right of the heterostructural setup and vanishes
elsewhere.

The thickness of the CM layer $n_{\rm CM}$ is varied from $0$ to $25$
layers, and we include up to $n_{\rm FM} = 500$ layers of FM in the
middle of the heterostructure. The vector components of the conical
exchange field are determined
by~\cite{Fritsch_NJP16_055005,Wu_PRB86_184517}
\begin{equation}
  \label{EqHelicalMagnet}
        {\bf h} = h_{0} \left\{ \cos \alpha {\bf y} + \sin \alpha
        \left[ \sin \left( \frac{\beta y}{a} \right) {\bf x} + \cos
          \left( \frac{\beta y}{a} \right) {\bf z} \right] \right\},
\end{equation}
with $h_{0} = 0.1$ fixing the exchange field in the CM layers, and
$a=1$ being the lattice constant. The opening angle $\alpha$ (measured
from $+y$ towards $+z$) and the turning angle $\beta$ (measured from
$+z$ towards $+x$) are fixed to the values of Ho, namely $\alpha =
80\,^{\circ}$ and $\beta = 30\,^{\circ}$.

\subsection{(Triplet) Pairing correlations}
\label{Sec2_2}

The superconducting pairing correlation between spins $\alpha$ and
$\beta$ can generally be evaluated as on-site average for times $t =
\tau$ and $t' = 0$ as
\begin{equation}
  \label{EqPairingCorrelationGeneral}
  f_{\alpha \beta}({\bf r}, \tau, 0) = \frac{1}{2}\bigl<
  \hat{\Psi}_{\alpha}({\bf r},\tau) \hat{\Psi}_{\beta}({\bf r},0)
  \bigr>,
\end{equation}
with $\hat{\Psi}_{\sigma}({\bf r},\tau)$ being the many-body field
operator for spin $\sigma$ at time $\tau$. The time-dependence is
introduced through the Heisenberg equation of motion. A pairing
correlation evaluated using Eq.~(\ref{EqPairingCorrelationGeneral}) is
local in space, leading to vanishing triplet contributions for $\tau =
0$ in accordance with the Pauli
principle.~\cite{Halterman_PRL99_127002} The nonvanishing
contributions for finite times $\tau$ are an example of odd-frequency
triplet pairing.~\cite{Bergeret_RMP77_1321} Substituting the field
operators valid for our setup and phase convention the spin-dependent
triplet pairing correlations read
\begin{widetext}
  \begin{equation}
    \label{EqTripletPairingCorrelations}
    \begin{split}
      f_{0}(y, \tau) & = \frac{1}{2} \bigl( f_{\uparrow \downarrow}(y,
      \tau) + f_{\downarrow \uparrow}(y, \tau) \bigr) = \frac{1}{2}
      \sum_{n} {\bigl( u_{n\uparrow}(y) v_{n\downarrow}^{*}(y) +
        u_{n\downarrow}(y) v_{n\uparrow}^{*}(y) \bigr)
        \zeta_{n}(\tau)} \\ f_{1}(y, \tau) & = \frac{1}{2} \bigl(
      f_{\uparrow \uparrow}(y, \tau) - f_{\downarrow \downarrow}(y,
      \tau) \bigr) = \frac{1}{2} \sum_{n} {\bigl( u_{n\uparrow}(y)
        v_{n\uparrow}^{*}(y) - u_{n\downarrow}(y)
        v_{n\downarrow}^{*}(y) \bigr) \zeta_{n}(\tau)}
    \end{split}
  \end{equation}
\end{widetext}
depending on position $y$ and time parameter $\tau$ (fixed to $\tau =
10$ throughout this work). The $\tau$ dependence is governed via
$\zeta_{n}(\tau)$ given by
\begin{equation}
  \label{EqTau}
  \zeta_{n}(\tau) = \cos (\varepsilon_{n} \tau) - i \sin
  (\varepsilon_{n} \tau) \bigl( 1 - 2 f(\varepsilon_{n}) \bigr).
\end{equation}
The different pairing correlations can be rewritten similarly to the
pairing matrix in Eq.~(\ref{EqPairingMatrix})
\begin{equation}
  \label{EqPairFunctionMatrix} \left( \begin{array}{cc}
    f_{\uparrow \uparrow} & f_{\uparrow \downarrow} \\ f_{\downarrow
      \uparrow} & f_{\downarrow \downarrow} \end{array} \right) =
  \left( f_{0} + {\hat {\bf \sigma}}{\bf f}\right) i {\hat \sigma}_{2}
  = \left( \begin{array}{cc} -f_{x} + i f_{y} & f_{0} + f_{z} \\ -
    f_{0} + f_{z} & f_{x} + i f_{y} \end{array} \right).
\end{equation}
In analogy to Eq.~(\ref{EqTripletPairingMatrix}) a restriction to the
vector components of the ${\bf f}$-vector yields the triplet pair
function matrix~\cite{Kawabata_JPSJ82_124702}
\begin{equation}
  \label{EqTripletPairFunctionMatrix}
        {\hat f} = \left( {\hat {\bf \sigma}}{\bf f}\right) i {\hat
          \sigma}_{2} = \left( \begin{array}{cc} -f_{x} + i f_{y} &
          f_{z} \\ f_{z} & f_{x} + i f_{y} \end{array} \right).
\end{equation}
Note, that we introduced an additional factor $i$ compared
to~\cite{Kawabata_JPSJ82_124702} to be consistent with the definition
of ${\hat \Delta}$ in Eq.~(\ref{EqTripletPairingMatrix}). Instead of
Eq.~(\ref{EqDDdagger}) we now have
\begin{equation}
  \label{EqFFdagger}
        {\hat f}{\hat f}^{\dagger} = |{\bf f}|^{2} {\hat \sigma}_{0} +
        i \left( {\bf f}\times{\bf f}^{*} \right) {\hat {\bf \sigma}},
\end{equation}
with the two analogues of $|{\bf d}|$ and ${\bf d}\times{\bf d}^{*}$,
being $|{\bf f}|$ and ${\bf f}\times{\bf f}^{*}$, being conveniently
expressed in terms of the ${\bf f}$-vector components
\begin{equation}
  \label{EqDVector}
  \begin{split}
    f_{x} & = \frac{1}{2} \left( -f_{\uparrow \uparrow} +
    f_{\downarrow \downarrow} \right) \\ f_{y} & = - \frac{i}{2}
    \left( f_{\uparrow \uparrow} + f_{\downarrow \downarrow} \right)
    \\ f_{z} & = \frac{1}{2} \left( f_{\uparrow \downarrow} +
    f_{\downarrow \uparrow} \right).
  \end{split}
\end{equation}

\section{Results and Discussion}
\label{Sec3}

\subsection{Influence of ferromagnet thickness $n_{\rm FM}$}
\label{Sec3_1}

\begin{figure*}
  \centering
  \includegraphics[width=0.6\textwidth,clip]{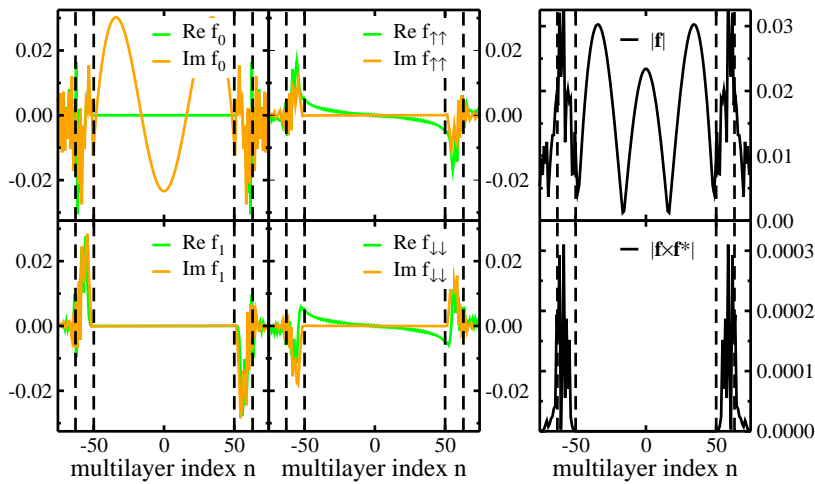}
  \caption{\label{Fig2} (colour online) Real (green) and imaginary
    (orange) parts of the spin-triplet pairing correlations $f_{0}$
    (upper left panel) and $f_{1}$ (lower left panel) according to
    Eq.~(\ref{EqTripletPairingCorrelations}). For better analysis the
    upper (lower) middle panels depict the real (green) and imaginary
    (orange) parts of $f_{\uparrow \uparrow}$ ($f_{\downarrow
      \downarrow}$) as components of $f_{1}$ shown in the lower left
    panel. Upper and lower right panels show the magnitude of the
    ${\bf f}$-vector and the magnitude of ${\bf f}\times{\bf f}^{*}$
    as introduced in Eq.~(\ref{EqFFdagger}), respectively. All data is
    shown depending on the multilayer index $n$, with $n = 0$ lying in
    the centre of the ferromagnetic layer. The vertical dashed lines
    indicate the FM/CM and CM/SC interfaces, respectively.}
\end{figure*}

The self-consistent calculation as described in Sec.~\ref{Sec2} yields
the spin-triplet pairing correlations $f_{0}$ and $f_{1}$ as defined
in Eq.~(\ref{EqTripletPairingCorrelations}). For an exemplary
heterostructure setup with $n_{\rm FM} = 100$ layers and one full turn
of the CM to either side, the upper (lower) left panels of
Fig.~\ref{Fig2} show results on the real (green) and imaginary
(orange) parts of the spin-triplet pairing correlations $f_{0}$
($f_{1}$), respectively.
The unequal-spin spin-triplet pairing correlation $f_{0}$ clearly
shows the expected oscillatory behaviour in the FM region of the
heterostructure. These oscillations change in the CM region and decay
in the adjacent SC layers. In contrast, the equal-spin spin-triplet
pairing correlations $f_{1}$ show pronounced features only inside the
CM region and decay into both the FM and the adjacent SC
layers. However, keeping in mind that $f_{1}$ has contributions from
$f_{\uparrow \uparrow}$ and $f_{\downarrow \downarrow}$ (see
Eq.~(\ref{EqTripletPairingCorrelations})), the upper (lower) middle
panels of Fig.~\ref{Fig2} show the real (green) and imaginary (orange)
parts of $f_{\uparrow \uparrow}$ ($f_{\downarrow \downarrow}$)
separately, clearly showing a nonvanishing contribution inside the FM
region for the single spin-channels. A more detailed analysis of the
influence of CM orientation in the heterostructure on the different
spin-triplet pairing correlations can be found in an earlier
work.~\cite{Fritsch_NJP16_055005} Recalling the pair function matrix
as of Eq.~(\ref{EqFFdagger}) the upper and lower right panels of
Fig.~\ref{Fig2} show the contributions to the magnitudes of the ${\bf
  f}$-vector and ${\bf f}\times{\bf f}^{*}$, respectively. The
magnitude of the ${\bf f}$-vector resembles the oscillatory patterns
already known from $f_{0}$ (upper left panel). The magnitude of ${\bf
  f}\times{\bf f}^{*}$, associated with the spin magnetic moment of
the Cooper pairs, shows leakage from the CM layer into the SC
layer. However, this leakage shows a strong decay and vanishes after a
few layers. On the contrary, there is no such leaking from the CM into
the FM region of the heterostructure.
\begin{figure*}
  \centering
  \includegraphics[width=0.75\textwidth,clip]{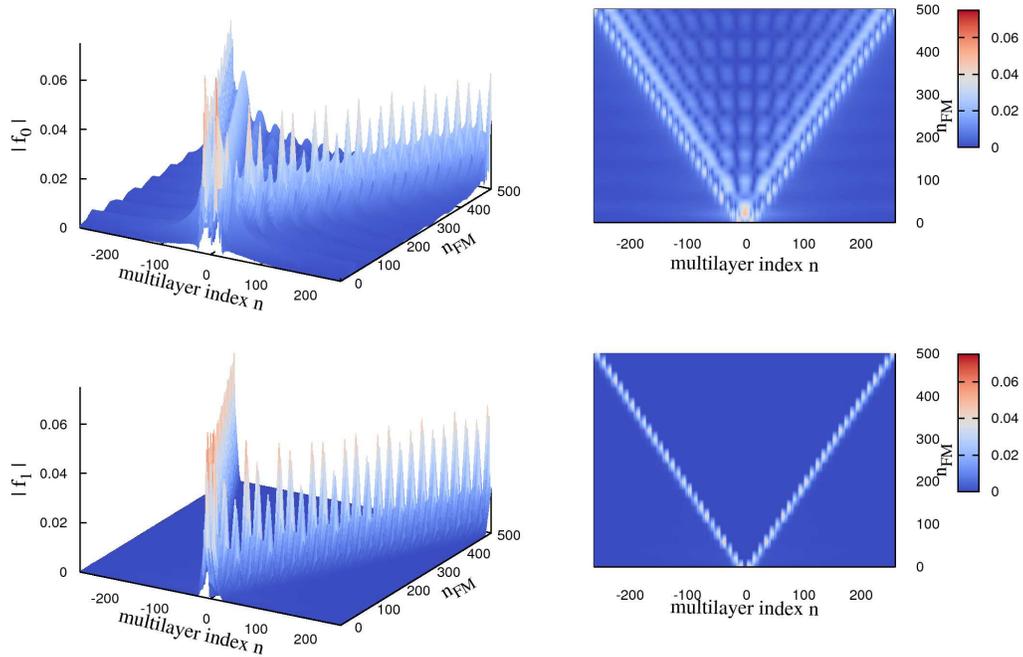}
  \caption{\label{Fig3} (colour online) Magnitudes of $f_{0}$ and
    $f_{1}$ depending on number of FM layers $n_{\rm FM}$ varied from
    $n_{\rm FM} = 0$ to $n_{\rm FM} = 500$ layers, respectively. Full
    data sets are shown in the left panels, whereas the right panels
    depict a top view of the data.}
\end{figure*}
Keeping the number of CM layers fixed to one full turn we now vary the
number of FM layers from $n_{\rm FM} = 0$ to $n_{\rm FM} = 500$
layers. The magnitudes of $f_{0}$ and $f_{1}$ depending on $n_{\rm
  FM}$ are depicted in the upper and lower panels of
Fig.~\ref{Fig3}. Similarly to the results presented in
Fig.~\ref{Fig2}, the upper panels in Fig.~\ref{Fig3}, depicting the
magnitudes of $f_{0}$, show a pronounced oscillatory behaviour within
the FM region up to $n_{\rm FM} = 500$ layers. There are also much
weaker oscillations visible inside the SC region depending on the
number of FM layers considered. As expected from the lower left panels
of Fig.~\ref{Fig2} there is no contribution to the magnitude of
$f_{1}$ in the SC or FM region of the heterostructure. The only sharp
signals originate from the CM regions and are of similar strength as
$f_{0}$. The strengths of both, $f_{0}$ and $f_{1}$, are influenced by
the number of FM layers $n_{\rm FM}$, especially visible for small
values of $n_{\rm FM}$, where oscillations are seen in $f_{0}$ and
$f_{1}$.
\begin{figure*}
  \centering
  \includegraphics[width=0.75\textwidth,clip]{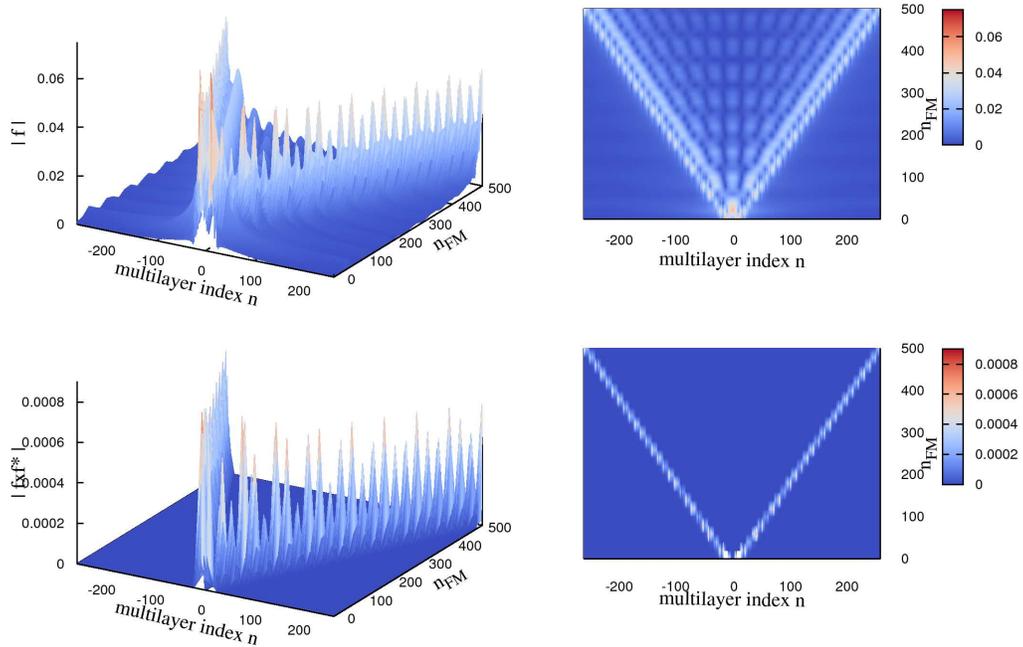}
  \caption{\label{Fig4} (colour online) Magnitudes of the ${\bf
      f}$-vector and ${\bf f}\times{\bf f}^{*}$ as introduced in
    Eq.~(\ref{EqFFdagger}) depending on the number of FM layers
    $n_{\rm FM}$ varied from $n_{\rm FM} = 0$ to $n_{\rm FM} = 500$
    layers, respectively. Full data sets are shown in the left panels,
    whereas the right panels depict a top view of the data.}
\end{figure*}
Recalling the pair function matrix as of Eq.~(\ref{EqFFdagger}) the
magnitudes of the ${\bf f}$-vector and ${\bf f}\times{\bf f}^{*}$ are
shown in the upper and lower panels of Fig.~\ref{Fig4},
respectively. As can be seen from the upper panels the magnitude of
the ${\bf f}$-vector is quite similar to the magnitude of $f_{0}$
already presented in Fig.~\ref{Fig3}. Keeping in mind the appearance
of the magnitude of ${\bf f}\times{\bf f}^{*}$ shown in the lower
right panel of Fig.~\ref{Fig2} there is no influence of the number of
FM layers $n_{\rm FM}$ on this equal-spin pairing correlation. For all
$n_{\rm FM}$ considered here, the strongest contributions to the
magnitude of ${\bf f}\times{\bf f}^{*}$ stem from the CM layers only,
with smaller and fast decaying contributions in the adjacent SC
layers. No contributions are visible inside the FM layers, as already
noted in the discussion of Fig.~\ref{Fig2}. To summarise this section,
the number of FM layers $n_{\rm FM}$ in the heterostructure has no
influence on the magnitudes of $f_{1}$ and ${\bf f}\times{\bf f}^{*}$,
whereas the number of maxima in the magnitudes of $f_{0}$ and the
${\bf f}$-vector are increased with increasing number of FM layers
$n_{\rm FM}$.

\subsection{Influence of conical magnet thickness $n_{\rm CM}$}
\label{Sec3_2}

\begin{figure*}
  \centering
  \includegraphics[width=0.95\textwidth,clip]{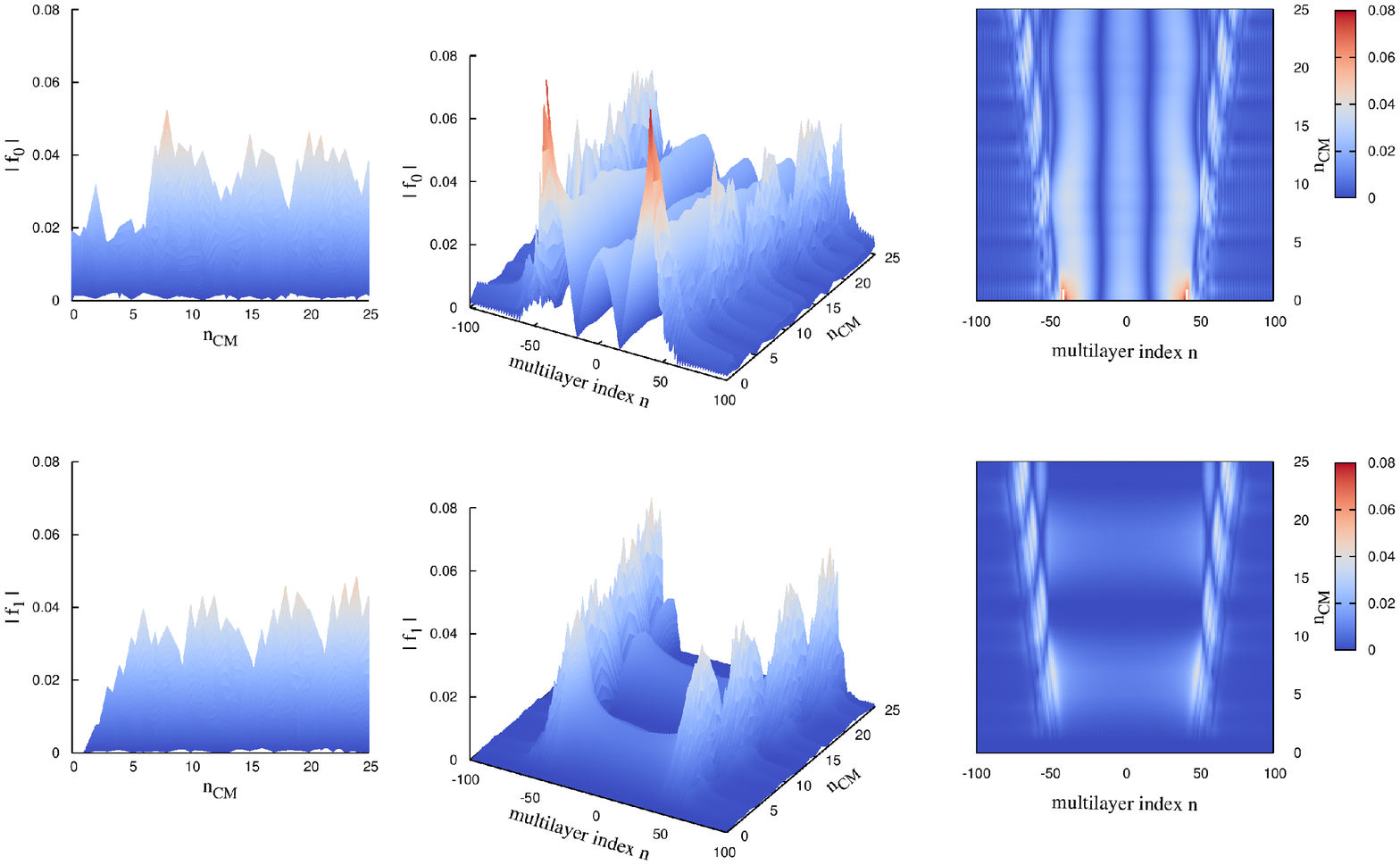}
  \caption{\label{Fig5} (colour online) Magnitudes of $f_{0}$ and
    $f_{1}$ depending on number of CM layers $n_{\rm CM}$ varied from
    $n_{\rm CM} = 0$ to $n_{\rm CM} = 25$ layers, respectively. Full
    data sets are shown in the middle panels, whereas in the left and
    right panels the same data is shown when viewed from the $yz$
    plane (right CM region only) and as a top view, respectively.}
\end{figure*}
Understanding the behaviour of FM layers on the various spin-triplet
pairing correlations as discussed in Sec.~\ref{Sec3_1}, we are now
interested in the influence of the number of CM layers $n_{\rm
  CM}$. For the following calculations we fix the number of FM layers
to $n_{\rm FM} = 100$ layers and vary $n_{\rm CM}$ from $0$ to $25$
layers, respectively. The magnitudes of $f_{0}$ and $f_{1}$ depending
on $n_{\rm CM}$ are depicted in the upper and lower panels of
Fig.~\ref{Fig5}. Keeping in mind that $n_{\rm FM}$ is fixed to $100$
the upper panels of Fig.~\ref{Fig5} show the already familiar
oscillatory patterns in the magnitude of $f_{0}$. In contrary to
previous results, the magnitude of $f_{0}$ is affected by the number
of CM layers $n_{\rm CM}$, as can be clearly seen in the intensity
variations in the FM region shown in the upper right panel of
Fig.~\ref{Fig5}. As the number of CM layers $n_{\rm CM}$ increase, the
magnitudes of both, $f_{0}$ and $f_{1}$, show a strongly oscillatory
behaviour in the CM region of the heterostructure. Minima appear at
multiples of half-integer turns of the full conical magnetic
structure, whereas the maxima lie inbetween. Also, there is a phase
shift present between the maxima appearing in the magnitudes of
$f_{0}$ (upper left panel of Fig.~\ref{Fig5}) and $f_{1}$ (lower left
panel of Fig.~\ref{Fig5}). In addition, the magnitude of $f_{1}$ show
nonvanishing contributions in the FM region of the heterostructure,
showing the same oscillatory dependence as in the CM regions,
respectively. Apparently, at least a quarter of a full conical
magnetic structure has to be present in the heterostructure to obtain
a sizable effect on the magnitude of $f_{1}$ (lower panels of
Fig.~\ref{Fig5}).
\begin{figure*}
  \centering
  \includegraphics[width=0.95\textwidth,clip]{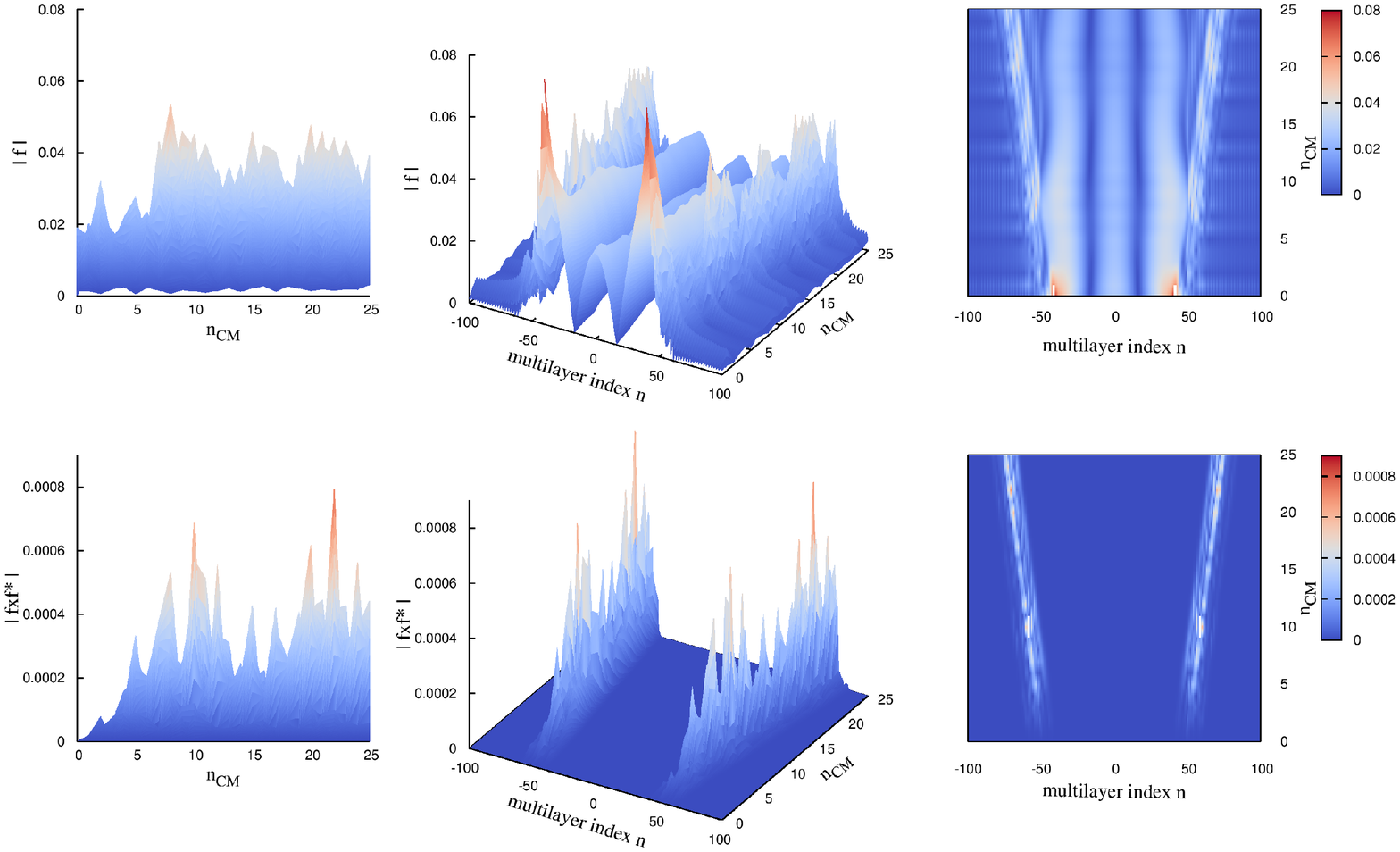}
  \caption{\label{Fig6} (colour online) Magnitudes of the ${\bf
      f}$-vector and ${\bf f}\times{\bf f}^{*}$ as introduced in
    Eq.~(\ref{EqFFdagger}) depending on the number of CM layers
    $n_{\rm CM}$ varied from $n_{\rm CM} = 0$ to $n_{\rm CM} = 25$
    layers, respectively. Full data sets are shown in the middle
    panels, whereas in the left and right panels the same data is
    shown when viewed from the $yz$ plane (right CM region only) and
    as a top view, respectively.}
\end{figure*}
The respective magnitudes of the ${\bf f}$-vector and ${\bf
  f}\times{\bf f}^{*}$ are shown in the upper and lower panels of
Fig.~\ref{Fig6}, respectively. With the magnitude of $f_{1}$ now
showing considerable contributions, the difference between the
magnitudes of $f_{0}$ (upper panels of Fig.~\ref{Fig5}) and the ${\bf
  f}$-vector (upper panels of Fig.~\ref{Fig6}) is larger compared to
Sec.~\ref{Sec3_1}. The magnitudes of $f_{1}$ are added to those of
$f_{0}$ to yield the magnitudes of the ${\bf f}$-vector, and are thus
responsible for the small intensity variations in the FM region (upper
right panel of Fig.~\ref{Fig6}). With the slight phase shift present
between the maxima appearing in the magnitudes of $f_{0}$ and $f_{1}$
the features visible in the magnitude of the ${\bf f}$-vector (lower
left panel of Fig.~\ref{Fig6}) are less sharp. From the lower panels
of Fig.~\ref{Fig6}, depicting the magnitudes of ${\bf f}\times{\bf
  f}^{*}$, we observe a strong influence of the number of CM layers
$n_{\rm CM}$ on the intensity.  This again shows oscillatory behaviour
related to the number of turns of the conical magnetic
structure. However, in this case there is no increase in the
penetration of ${\bf f}\times{\bf f}^{*}$ into the FM or SC
region. Summarising this section, the number of CM layers $n_{\rm CM}$
has a more pronounced effect on the spin-triplet pairing correlations
compared to the number of FM layers $n_{\rm FM}$ as discussed in
Sec.~\ref{Sec3_1}. The oscillations in the magnitudes of $f_{1}$ shown
in the FM and CM regions of the heterostructure are also visible in
the magnitude of the ${\bf f}$-vector. In addition, the first sizable
contributions to the magnitudes of $f_{1}$ and ${\bf f}\times{\bf
  f}^{*}$ require at least a number of CM layers corresponding to a
quarter of a full conical magnetic turn.

\section{Summary and Outlook}
\label{SummaryAndOutlook}

In summary, we presented results on the spin-triplet pairing
correlations in a SC/FM/SC heterostructure including Ho as a spin-flip
source at the interfaces. The calculations were based on
self-consistent solutions of the microscopic BdG equations in the
clean limit. In addition to the spin-triplet pairing correlations
$f_{0}$ and $f_{1}$, the magnitudes of the ${\bf f}$-vector and ${\bf
  f}\times{\bf f}^{*}$ have also been calculated. These allow for a
deeper understanding of the influence of varying thicknesses of the FM
and CM layers on the total and equal-spin spin-triplet correlations,
respectively. The leaking of spin magnetic moment of the Cooper pairs
into the SC region of the heterostructure is limited and decays
rapidly after only a few layers of SC. This behaviour is not
influenced by the thickness of FM or CM layers. On the contrary,
sizeable contributions to the magnitude of $f_{1}$ require a CM
thickness corresponding to at least a quarter of the full magnetic
cone structure. This is in agreement with experimental
observations~\cite{Robinson_Science329_59} where similar
heterostructures have been investigated. Thereby, spin-triplet super
currents exhibit peak values at CM thicknesses corresponding to
roughly half a turn of the conical magnetic structure.

\section*{Acknowledgement}
This work has been financially supported by the EPSRC (EP/I037598/1)
and made use of computational resources of the University of Bristol.

%


\begin{thebibliography}{32}%
\makeatletter
\providecommand \@ifxundefined [1]{%
 \@ifx{#1\undefined}
}%
\providecommand \@ifnum [1]{%
 \ifnum #1\expandafter \@firstoftwo
 \else \expandafter \@secondoftwo
 \fi
}%
\providecommand \@ifx [1]{%
 \ifx #1\expandafter \@firstoftwo
 \else \expandafter \@secondoftwo
 \fi
}%
\providecommand \natexlab [1]{#1}%
\providecommand \enquote  [1]{``#1''}%
\providecommand \bibnamefont  [1]{#1}%
\providecommand \bibfnamefont [1]{#1}%
\providecommand \citenamefont [1]{#1}%
\providecommand \href@noop [0]{\@secondoftwo}%
\providecommand \href [0]{\begingroup \@sanitize@url \@href}%
\providecommand \@href[1]{\@@startlink{#1}\@@href}%
\providecommand \@@href[1]{\endgroup#1\@@endlink}%
\providecommand \@sanitize@url [0]{\catcode `\\12\catcode `\$12\catcode
  `\&12\catcode `\#12\catcode `\^12\catcode `\_12\catcode `\%12\relax}%
\providecommand \@@startlink[1]{}%
\providecommand \@@endlink[0]{}%
\providecommand \url  [0]{\begingroup\@sanitize@url \@url }%
\providecommand \@url [1]{\endgroup\@href {#1}{\urlprefix }}%
\providecommand \urlprefix  [0]{URL }%
\providecommand \Eprint [0]{\href }%
\providecommand \doibase [0]{http://dx.doi.org/}%
\providecommand \selectlanguage [0]{\@gobble}%
\providecommand \bibinfo  [0]{\@secondoftwo}%
\providecommand \bibfield  [0]{\@secondoftwo}%
\providecommand \translation [1]{[#1]}%
\providecommand \BibitemOpen [0]{}%
\providecommand \bibitemStop [0]{}%
\providecommand \bibitemNoStop [0]{.\EOS\space}%
\providecommand \EOS [0]{\spacefactor3000\relax}%
\providecommand \BibitemShut  [1]{\csname bibitem#1\endcsname}%
\let\auto@bib@innerbib\@empty
\bibitem [{\citenamefont {Fulde}\ and\ \citenamefont
  {Ferrell}(1964)}]{Fulde_PhysRev135_A550}%
  \BibitemOpen
  \bibfield  {author} {\bibinfo {author} {\bibfnamefont {P.}~\bibnamefont
  {Fulde}}\ and\ \bibinfo {author} {\bibfnamefont {R.~A.}\ \bibnamefont
  {Ferrell}},\ }\href@noop {} {\bibfield  {journal} {\bibinfo  {journal} {Phys.
  Rev.}\ }\textbf {\bibinfo {volume} {135}},\ \bibinfo {pages} {A550} (\bibinfo
  {year} {1964})}\BibitemShut {NoStop}%
\bibitem [{\citenamefont {Larkin}\ and\ \citenamefont
  {Ovchinnikov}(1965)}]{Larkin_JETP20_762}%
  \BibitemOpen
  \bibfield  {author} {\bibinfo {author} {\bibfnamefont {A.~I.}\ \bibnamefont
  {Larkin}}\ and\ \bibinfo {author} {\bibfnamefont {Y.~N.}\ \bibnamefont
  {Ovchinnikov}},\ }\href@noop {} {\bibfield  {journal} {\bibinfo  {journal}
  {Sov. Phys. JETP}\ }\textbf {\bibinfo {volume} {20}},\ \bibinfo {pages} {762}
  (\bibinfo {year} {1965})}\BibitemShut {NoStop}%
\bibitem [{\citenamefont {Bergeret}\ \emph {et~al.}(2001)\citenamefont
  {Bergeret}, \citenamefont {Volkov},\ and\ \citenamefont
  {Efetov}}]{Bergeret_PRL86_4096}%
  \BibitemOpen
  \bibfield  {author} {\bibinfo {author} {\bibfnamefont {F.~S.}\ \bibnamefont
  {Bergeret}}, \bibinfo {author} {\bibfnamefont {A.~F.}\ \bibnamefont
  {Volkov}}, \ and\ \bibinfo {author} {\bibfnamefont {K.~B.}\ \bibnamefont
  {Efetov}},\ }\href@noop {} {\bibfield  {journal} {\bibinfo  {journal} {Phys.
  Rev. Lett.}\ }\textbf {\bibinfo {volume} {86}},\ \bibinfo {pages} {4096}
  (\bibinfo {year} {2001})}\BibitemShut {NoStop}%
\bibitem [{\citenamefont {Buzdin}(2005)}]{Buzdin_RMP77_935}%
  \BibitemOpen
  \bibfield  {author} {\bibinfo {author} {\bibfnamefont {A.~I.}\ \bibnamefont
  {Buzdin}},\ }\href@noop {} {\bibfield  {journal} {\bibinfo  {journal} {Rev.
  Mod. Phys.}\ }\textbf {\bibinfo {volume} {77}},\ \bibinfo {pages} {935}
  (\bibinfo {year} {2005})}\BibitemShut {NoStop}%
\bibitem [{\citenamefont {Bergeret}\ \emph {et~al.}(2005)\citenamefont
  {Bergeret}, \citenamefont {Volkov},\ and\ \citenamefont
  {Efetov}}]{Bergeret_RMP77_1321}%
  \BibitemOpen
  \bibfield  {author} {\bibinfo {author} {\bibfnamefont {F.~S.}\ \bibnamefont
  {Bergeret}}, \bibinfo {author} {\bibfnamefont {A.~F.}\ \bibnamefont
  {Volkov}}, \ and\ \bibinfo {author} {\bibfnamefont {K.~B.}\ \bibnamefont
  {Efetov}},\ }\href@noop {} {\bibfield  {journal} {\bibinfo  {journal} {Rev.
  Mod. Phys.}\ }\textbf {\bibinfo {volume} {77}},\ \bibinfo {pages} {1321}
  (\bibinfo {year} {2005})}\BibitemShut {NoStop}%
\bibitem [{\citenamefont {Keizer}\ \emph {et~al.}(2006)\citenamefont {Keizer},
  \citenamefont {Goennenwein}, \citenamefont {Klapwijk}, \citenamefont {Miao},
  \citenamefont {Xiao},\ and\ \citenamefont {Gupta}}]{Keizer_Nature439_825}%
  \BibitemOpen
  \bibfield  {author} {\bibinfo {author} {\bibfnamefont {R.~S.}\ \bibnamefont
  {Keizer}}, \bibinfo {author} {\bibfnamefont {S.~T.~B.}\ \bibnamefont
  {Goennenwein}}, \bibinfo {author} {\bibfnamefont {T.~M.}\ \bibnamefont
  {Klapwijk}}, \bibinfo {author} {\bibfnamefont {G.}~\bibnamefont {Miao}},
  \bibinfo {author} {\bibfnamefont {G.}~\bibnamefont {Xiao}}, \ and\ \bibinfo
  {author} {\bibfnamefont {A.}~\bibnamefont {Gupta}},\ }\href@noop {}
  {\bibfield  {journal} {\bibinfo  {journal} {Nature}\ }\textbf {\bibinfo
  {volume} {439}},\ \bibinfo {pages} {825} (\bibinfo {year}
  {2006})}\BibitemShut {NoStop}%
\bibitem [{\citenamefont {Anwar}\ \emph {et~al.}(2010)\citenamefont {Anwar},
  \citenamefont {Czeschka}, \citenamefont {Hesselberth}, \citenamefont
  {Porcu},\ and\ \citenamefont {Aarts}}]{Anwar_PRB82_100501}%
  \BibitemOpen
  \bibfield  {author} {\bibinfo {author} {\bibfnamefont {M.~S.}\ \bibnamefont
  {Anwar}}, \bibinfo {author} {\bibfnamefont {F.}~\bibnamefont {Czeschka}},
  \bibinfo {author} {\bibfnamefont {M.}~\bibnamefont {Hesselberth}}, \bibinfo
  {author} {\bibfnamefont {M.}~\bibnamefont {Porcu}}, \ and\ \bibinfo {author}
  {\bibfnamefont {J.}~\bibnamefont {Aarts}},\ }\href@noop {} {\bibfield
  {journal} {\bibinfo  {journal} {Phys. Rev. B}\ }\textbf {\bibinfo {volume}
  {82}},\ \bibinfo {pages} {100501} (\bibinfo {year} {2010})}\BibitemShut
  {NoStop}%
\bibitem [{\citenamefont {Visani}\ \emph {et~al.}(2012)\citenamefont {Visani},
  \citenamefont {Sefrioui}, \citenamefont {Tornos}, \citenamefont {Leon},
  \citenamefont {Briatico}, \citenamefont {Bibes}, \citenamefont
  {Barth\'{e}l\'{e}my}, \citenamefont {Santamar\'{i}a},\ and\ \citenamefont
  {Villegas}}]{Visani_NatPhys8_539}%
  \BibitemOpen
  \bibfield  {author} {\bibinfo {author} {\bibfnamefont {C.}~\bibnamefont
  {Visani}}, \bibinfo {author} {\bibfnamefont {Z.}~\bibnamefont {Sefrioui}},
  \bibinfo {author} {\bibfnamefont {J.}~\bibnamefont {Tornos}}, \bibinfo
  {author} {\bibfnamefont {C.}~\bibnamefont {Leon}}, \bibinfo {author}
  {\bibfnamefont {J.}~\bibnamefont {Briatico}}, \bibinfo {author}
  {\bibfnamefont {M.}~\bibnamefont {Bibes}}, \bibinfo {author} {\bibfnamefont
  {A.}~\bibnamefont {Barth\'{e}l\'{e}my}}, \bibinfo {author} {\bibfnamefont
  {J.}~\bibnamefont {Santamar\'{i}a}}, \ and\ \bibinfo {author} {\bibfnamefont
  {J.~E.}\ \bibnamefont {Villegas}},\ }\href@noop {} {\bibfield  {journal}
  {\bibinfo  {journal} {Nature Phys.}\ }\textbf {\bibinfo {volume} {8}},\
  \bibinfo {pages} {539} (\bibinfo {year} {2012})}\BibitemShut {NoStop}%
\bibitem [{\citenamefont {Khaire}\ \emph {et~al.}(2010)\citenamefont {Khaire},
  \citenamefont {Khasawneh}, \citenamefont {{Pratt Jr.}},\ and\ \citenamefont
  {Birge}}]{Khaire_PRL104_137002}%
  \BibitemOpen
  \bibfield  {author} {\bibinfo {author} {\bibfnamefont {T.~S.}\ \bibnamefont
  {Khaire}}, \bibinfo {author} {\bibfnamefont {M.~A.}\ \bibnamefont
  {Khasawneh}}, \bibinfo {author} {\bibfnamefont {W.~P.}\ \bibnamefont {{Pratt
  Jr.}}}, \ and\ \bibinfo {author} {\bibfnamefont {N.~O.}\ \bibnamefont
  {Birge}},\ }\href@noop {} {\bibfield  {journal} {\bibinfo  {journal} {Phys.
  Rev. Lett.}\ }\textbf {\bibinfo {volume} {104}},\ \bibinfo {pages} {137002}
  (\bibinfo {year} {2010})}\BibitemShut {NoStop}%
\bibitem [{\citenamefont {Klose}\ \emph {et~al.}(2012)\citenamefont {Klose},
  \citenamefont {Khaire}, \citenamefont {Wang}, \citenamefont {{Pratt Jr.}},
  \citenamefont {Birge}, \citenamefont {Mc{M}orran}, \citenamefont {Ginley},
  \citenamefont {Borchers}, \citenamefont {Kirby}, \citenamefont {Maranville},\
  and\ \citenamefont {Unguris}}]{Klose_PRL108_127002}%
  \BibitemOpen
  \bibfield  {author} {\bibinfo {author} {\bibfnamefont {C.}~\bibnamefont
  {Klose}}, \bibinfo {author} {\bibfnamefont {T.~S.}\ \bibnamefont {Khaire}},
  \bibinfo {author} {\bibfnamefont {Y.}~\bibnamefont {Wang}}, \bibinfo {author}
  {\bibfnamefont {W.~P.}\ \bibnamefont {{Pratt Jr.}}}, \bibinfo {author}
  {\bibfnamefont {N.~O.}\ \bibnamefont {Birge}}, \bibinfo {author}
  {\bibfnamefont {B.~J.}\ \bibnamefont {Mc{M}orran}}, \bibinfo {author}
  {\bibfnamefont {T.~P.}\ \bibnamefont {Ginley}}, \bibinfo {author}
  {\bibfnamefont {J.~A.}\ \bibnamefont {Borchers}}, \bibinfo {author}
  {\bibfnamefont {B.~J.}\ \bibnamefont {Kirby}}, \bibinfo {author}
  {\bibfnamefont {B.~B.}\ \bibnamefont {Maranville}}, \ and\ \bibinfo {author}
  {\bibfnamefont {J.}~\bibnamefont {Unguris}},\ }\href@noop {} {\bibfield
  {journal} {\bibinfo  {journal} {Phys. Rev. Lett.}\ }\textbf {\bibinfo
  {volume} {108}},\ \bibinfo {pages} {127002} (\bibinfo {year}
  {2012})}\BibitemShut {NoStop}%
\bibitem [{\citenamefont {Gingrich}\ \emph {et~al.}(2012)\citenamefont
  {Gingrich}, \citenamefont {Quarterman}, \citenamefont {Wang}, \citenamefont
  {Loloee}, \citenamefont {{Pratt, Jr.}},\ and\ \citenamefont
  {Birge}}]{Gingrich_PRB86_224506}%
  \BibitemOpen
  \bibfield  {author} {\bibinfo {author} {\bibfnamefont {E.~C.}\ \bibnamefont
  {Gingrich}}, \bibinfo {author} {\bibfnamefont {P.}~\bibnamefont
  {Quarterman}}, \bibinfo {author} {\bibfnamefont {Y.}~\bibnamefont {Wang}},
  \bibinfo {author} {\bibfnamefont {R.}~\bibnamefont {Loloee}}, \bibinfo
  {author} {\bibfnamefont {W.~P.}\ \bibnamefont {{Pratt, Jr.}}}, \ and\
  \bibinfo {author} {\bibfnamefont {N.~O.}\ \bibnamefont {Birge}},\ }\href@noop
  {} {\bibfield  {journal} {\bibinfo  {journal} {Phys. Rev. B}\ }\textbf
  {\bibinfo {volume} {86}},\ \bibinfo {pages} {224506} (\bibinfo {year}
  {2012})}\BibitemShut {NoStop}%
\bibitem [{\citenamefont {Zdravkov}\ \emph {et~al.}(2013)\citenamefont
  {Zdravkov}, \citenamefont {Kehrle}, \citenamefont {Obermeier}, \citenamefont
  {Lenk}, \citenamefont {{Krug von Nidda}}, \citenamefont {M\"{u}ller},
  \citenamefont {Kupriyanov}, \citenamefont {Sidorenko}, \citenamefont {Horn},
  \citenamefont {Tidecks},\ and\ \citenamefont
  {Tagirov}}]{Zdravkov_PRB87_144507}%
  \BibitemOpen
  \bibfield  {author} {\bibinfo {author} {\bibfnamefont {V.~I.}\ \bibnamefont
  {Zdravkov}}, \bibinfo {author} {\bibfnamefont {J.}~\bibnamefont {Kehrle}},
  \bibinfo {author} {\bibfnamefont {G.}~\bibnamefont {Obermeier}}, \bibinfo
  {author} {\bibfnamefont {D.}~\bibnamefont {Lenk}}, \bibinfo {author}
  {\bibfnamefont {H.-A.}\ \bibnamefont {{Krug von Nidda}}}, \bibinfo {author}
  {\bibfnamefont {C.}~\bibnamefont {M\"{u}ller}}, \bibinfo {author}
  {\bibfnamefont {M.~Y.}\ \bibnamefont {Kupriyanov}}, \bibinfo {author}
  {\bibfnamefont {A.~S.}\ \bibnamefont {Sidorenko}}, \bibinfo {author}
  {\bibfnamefont {S.}~\bibnamefont {Horn}}, \bibinfo {author} {\bibfnamefont
  {R.}~\bibnamefont {Tidecks}}, \ and\ \bibinfo {author} {\bibfnamefont
  {L.~R.}\ \bibnamefont {Tagirov}},\ }\href@noop {} {\bibfield  {journal}
  {\bibinfo  {journal} {Phys. Rev. B}\ }\textbf {\bibinfo {volume} {87}},\
  \bibinfo {pages} {144507} (\bibinfo {year} {2013})}\BibitemShut {NoStop}%
\bibitem [{\citenamefont {Hal\'{a}sz}\ \emph {et~al.}(2011)\citenamefont
  {Hal\'{a}sz}, \citenamefont {Blamire},\ and\ \citenamefont
  {Robinson}}]{Halasz_PRB84_024517}%
  \BibitemOpen
  \bibfield  {author} {\bibinfo {author} {\bibfnamefont {G.~B.}\ \bibnamefont
  {Hal\'{a}sz}}, \bibinfo {author} {\bibfnamefont {M.~G.}\ \bibnamefont
  {Blamire}}, \ and\ \bibinfo {author} {\bibfnamefont {J.~W.~A.}\ \bibnamefont
  {Robinson}},\ }\href@noop {} {\bibfield  {journal} {\bibinfo  {journal}
  {Phys. Rev. B}\ }\textbf {\bibinfo {volume} {84}},\ \bibinfo {pages} {024517}
  (\bibinfo {year} {2011})}\BibitemShut {NoStop}%
\bibitem [{\citenamefont {Sosnin}\ \emph {et~al.}(2006)\citenamefont {Sosnin},
  \citenamefont {Cho}, \citenamefont {Petrashov},\ and\ \citenamefont
  {Volkov}}]{Sosnin_PRL96_157002}%
  \BibitemOpen
  \bibfield  {author} {\bibinfo {author} {\bibfnamefont {I.}~\bibnamefont
  {Sosnin}}, \bibinfo {author} {\bibfnamefont {H.}~\bibnamefont {Cho}},
  \bibinfo {author} {\bibfnamefont {V.~T.}\ \bibnamefont {Petrashov}}, \ and\
  \bibinfo {author} {\bibfnamefont {A.~F.}\ \bibnamefont {Volkov}},\
  }\href@noop {} {\bibfield  {journal} {\bibinfo  {journal} {Phys. Rev. Lett.}\
  }\textbf {\bibinfo {volume} {96}},\ \bibinfo {pages} {157002} (\bibinfo
  {year} {2006})}\BibitemShut {NoStop}%
\bibitem [{\citenamefont {Hal\'{a}sz}\ \emph {et~al.}(2009)\citenamefont
  {Hal\'{a}sz}, \citenamefont {Robinson}, \citenamefont {Annett},\ and\
  \citenamefont {Blamire}}]{Halasz_PRB79_224505}%
  \BibitemOpen
  \bibfield  {author} {\bibinfo {author} {\bibfnamefont {G.~B.}\ \bibnamefont
  {Hal\'{a}sz}}, \bibinfo {author} {\bibfnamefont {J.~W.~A.}\ \bibnamefont
  {Robinson}}, \bibinfo {author} {\bibfnamefont {J.~F.}\ \bibnamefont
  {Annett}}, \ and\ \bibinfo {author} {\bibfnamefont {M.~G.}\ \bibnamefont
  {Blamire}},\ }\href@noop {} {\bibfield  {journal} {\bibinfo  {journal} {Phys.
  Rev. B}\ }\textbf {\bibinfo {volume} {79}},\ \bibinfo {pages} {224505}
  (\bibinfo {year} {2009})}\BibitemShut {NoStop}%
\bibitem [{\citenamefont {Robinson}\ \emph {et~al.}(2010)\citenamefont
  {Robinson}, \citenamefont {Witt},\ and\ \citenamefont
  {Blamire}}]{Robinson_Science329_59}%
  \BibitemOpen
  \bibfield  {author} {\bibinfo {author} {\bibfnamefont {J.~W.~A.}\
  \bibnamefont {Robinson}}, \bibinfo {author} {\bibfnamefont {J.~D.~S.}\
  \bibnamefont {Witt}}, \ and\ \bibinfo {author} {\bibfnamefont {M.~G.}\
  \bibnamefont {Blamire}},\ }\href@noop {} {\bibfield  {journal} {\bibinfo
  {journal} {Science}\ }\textbf {\bibinfo {volume} {329}},\ \bibinfo {pages}
  {59} (\bibinfo {year} {2010})}\BibitemShut {NoStop}%
\bibitem [{\citenamefont {Eschrig}\ \emph {et~al.}(2003)\citenamefont
  {Eschrig}, \citenamefont {Kopu}, \citenamefont {Cuevas},\ and\ \citenamefont
  {Sch\"{o}n}}]{Eschrig_PRL90_137003}%
  \BibitemOpen
  \bibfield  {author} {\bibinfo {author} {\bibfnamefont {M.}~\bibnamefont
  {Eschrig}}, \bibinfo {author} {\bibfnamefont {J.}~\bibnamefont {Kopu}},
  \bibinfo {author} {\bibfnamefont {J.~C.}\ \bibnamefont {Cuevas}}, \ and\
  \bibinfo {author} {\bibfnamefont {G.}~\bibnamefont {Sch\"{o}n}},\ }\href@noop
  {} {\bibfield  {journal} {\bibinfo  {journal} {Phys. Rev. Lett.}\ }\textbf
  {\bibinfo {volume} {90}},\ \bibinfo {pages} {137003} (\bibinfo {year}
  {2003})}\BibitemShut {NoStop}%
\bibitem [{\citenamefont {Eschrig}\ and\ \citenamefont
  {L\"{o}fwander}(2008)}]{Eschrig_NaturePhys4_138}%
  \BibitemOpen
  \bibfield  {author} {\bibinfo {author} {\bibfnamefont {M.}~\bibnamefont
  {Eschrig}}\ and\ \bibinfo {author} {\bibfnamefont {T.}~\bibnamefont
  {L\"{o}fwander}},\ }\href@noop {} {\bibfield  {journal} {\bibinfo  {journal}
  {Nature Phys.}\ }\textbf {\bibinfo {volume} {4}},\ \bibinfo {pages} {138}
  (\bibinfo {year} {2008})}\BibitemShut {NoStop}%
\bibitem [{\citenamefont {Alidoust}\ \emph {et~al.}(2010)\citenamefont
  {Alidoust}, \citenamefont {Linder}, \citenamefont {Rashedi}, \citenamefont
  {Yokoyama},\ and\ \citenamefont {Sudb\o}}]{Alidoust_PRB81_014512}%
  \BibitemOpen
  \bibfield  {author} {\bibinfo {author} {\bibfnamefont {M.}~\bibnamefont
  {Alidoust}}, \bibinfo {author} {\bibfnamefont {J.}~\bibnamefont {Linder}},
  \bibinfo {author} {\bibfnamefont {G.}~\bibnamefont {Rashedi}}, \bibinfo
  {author} {\bibfnamefont {T.}~\bibnamefont {Yokoyama}}, \ and\ \bibinfo
  {author} {\bibfnamefont {A.}~\bibnamefont {Sudb\o}},\ }\href@noop {}
  {\bibfield  {journal} {\bibinfo  {journal} {Phys. Rev. B}\ }\textbf {\bibinfo
  {volume} {81}},\ \bibinfo {pages} {014512} (\bibinfo {year}
  {2010})}\BibitemShut {NoStop}%
\bibitem [{\citenamefont {Volkov}\ \emph {et~al.}(2003)\citenamefont {Volkov},
  \citenamefont {Bergeret},\ and\ \citenamefont
  {Efetov}}]{Volkov_PRL90_117006}%
  \BibitemOpen
  \bibfield  {author} {\bibinfo {author} {\bibfnamefont {A.~F.}\ \bibnamefont
  {Volkov}}, \bibinfo {author} {\bibfnamefont {F.~S.}\ \bibnamefont
  {Bergeret}}, \ and\ \bibinfo {author} {\bibfnamefont {K.~B.}\ \bibnamefont
  {Efetov}},\ }\href@noop {} {\bibfield  {journal} {\bibinfo  {journal} {Phys.
  Rev. Lett.}\ }\textbf {\bibinfo {volume} {90}},\ \bibinfo {pages} {117006}
  (\bibinfo {year} {2003})}\BibitemShut {NoStop}%
\bibitem [{\citenamefont {Wu}\ \emph {et~al.}(2012)\citenamefont {Wu},
  \citenamefont {Valls},\ and\ \citenamefont {Halterman}}]{Wu_PRB86_184517}%
  \BibitemOpen
  \bibfield  {author} {\bibinfo {author} {\bibfnamefont {C.-T.}\ \bibnamefont
  {Wu}}, \bibinfo {author} {\bibfnamefont {O.~T.}\ \bibnamefont {Valls}}, \
  and\ \bibinfo {author} {\bibfnamefont {K.}~\bibnamefont {Halterman}},\
  }\href@noop {} {\bibfield  {journal} {\bibinfo  {journal} {Phys. Rev. B}\
  }\textbf {\bibinfo {volume} {86}},\ \bibinfo {pages} {184517} (\bibinfo
  {year} {2012})}\BibitemShut {NoStop}%
\bibitem [{\citenamefont {Fritsch}\ and\ \citenamefont
  {Annett}(2014{\natexlab{a}})}]{Fritsch_NJP16_055005}%
  \BibitemOpen
  \bibfield  {author} {\bibinfo {author} {\bibfnamefont {D.}~\bibnamefont
  {Fritsch}}\ and\ \bibinfo {author} {\bibfnamefont {J.~F.}\ \bibnamefont
  {Annett}},\ }\href@noop {} {\bibfield  {journal} {\bibinfo  {journal} {New J.
  Phys.}\ }\textbf {\bibinfo {volume} {16}},\ \bibinfo {pages} {055005}
  (\bibinfo {year} {2014}{\natexlab{a}})}\BibitemShut {NoStop}%
\bibitem [{\citenamefont {Fritsch}\ and\ \citenamefont
  {Annett}(2014{\natexlab{b}})}]{Fritsch_JPCM26_274212}%
  \BibitemOpen
  \bibfield  {author} {\bibinfo {author} {\bibfnamefont {D.}~\bibnamefont
  {Fritsch}}\ and\ \bibinfo {author} {\bibfnamefont {J.~F.}\ \bibnamefont
  {Annett}},\ }\href@noop {} {\bibfield  {journal} {\bibinfo  {journal} {J.
  Phys.: Condens. Matter}\ }\textbf {\bibinfo {volume} {26}},\ \bibinfo {pages}
  {274212} (\bibinfo {year} {2014}{\natexlab{b}})}\BibitemShut {NoStop}%
\bibitem [{\citenamefont {Fominov}\ \emph {et~al.}(2007)\citenamefont
  {Fominov}, \citenamefont {Volkov},\ and\ \citenamefont
  {Efetov}}]{Fominov_PRB75_104509}%
  \BibitemOpen
  \bibfield  {author} {\bibinfo {author} {\bibfnamefont {Y.~V.}\ \bibnamefont
  {Fominov}}, \bibinfo {author} {\bibfnamefont {A.~F.}\ \bibnamefont {Volkov}},
  \ and\ \bibinfo {author} {\bibfnamefont {K.~B.}\ \bibnamefont {Efetov}},\
  }\href@noop {} {\bibfield  {journal} {\bibinfo  {journal} {Phys. Rev. B}\
  }\textbf {\bibinfo {volume} {75}},\ \bibinfo {pages} {104509} (\bibinfo
  {year} {2007})}\BibitemShut {NoStop}%
\bibitem [{\citenamefont {Kawabata}\ \emph {et~al.}(2013)\citenamefont
  {Kawabata}, \citenamefont {Asano}, \citenamefont {Tanaka},\ and\
  \citenamefont {Golubov}}]{Kawabata_JPSJ82_124702}%
  \BibitemOpen
  \bibfield  {author} {\bibinfo {author} {\bibfnamefont {S.}~\bibnamefont
  {Kawabata}}, \bibinfo {author} {\bibfnamefont {Y.}~\bibnamefont {Asano}},
  \bibinfo {author} {\bibfnamefont {Y.}~\bibnamefont {Tanaka}}, \ and\ \bibinfo
  {author} {\bibfnamefont {A.~A.}\ \bibnamefont {Golubov}},\ }\href@noop {}
  {\bibfield  {journal} {\bibinfo  {journal} {J. Phys. Soc. Jpn.}\ }\textbf
  {\bibinfo {volume} {82}},\ \bibinfo {pages} {124702} (\bibinfo {year}
  {2013})}\BibitemShut {NoStop}%
\bibitem [{\citenamefont {Halterman}\ \emph {et~al.}(2007)\citenamefont
  {Halterman}, \citenamefont {Barsic},\ and\ \citenamefont
  {Valls}}]{Halterman_PRL99_127002}%
  \BibitemOpen
  \bibfield  {author} {\bibinfo {author} {\bibfnamefont {K.}~\bibnamefont
  {Halterman}}, \bibinfo {author} {\bibfnamefont {P.~H.}\ \bibnamefont
  {Barsic}}, \ and\ \bibinfo {author} {\bibfnamefont {O.~T.}\ \bibnamefont
  {Valls}},\ }\href@noop {} {\bibfield  {journal} {\bibinfo  {journal} {Phys.
  Rev. Lett.}\ }\textbf {\bibinfo {volume} {99}},\ \bibinfo {pages} {127002}
  (\bibinfo {year} {2007})}\BibitemShut {NoStop}%
\bibitem [{\citenamefont {Halterman}\ \emph {et~al.}(2008)\citenamefont
  {Halterman}, \citenamefont {Valls},\ and\ \citenamefont
  {Barsic}}]{Halterman_PRB77_174511}%
  \BibitemOpen
  \bibfield  {author} {\bibinfo {author} {\bibfnamefont {K.}~\bibnamefont
  {Halterman}}, \bibinfo {author} {\bibfnamefont {O.~T.}\ \bibnamefont
  {Valls}}, \ and\ \bibinfo {author} {\bibfnamefont {P.~H.}\ \bibnamefont
  {Barsic}},\ }\href@noop {} {\bibfield  {journal} {\bibinfo  {journal} {Phys.
  Rev. B}\ }\textbf {\bibinfo {volume} {77}},\ \bibinfo {pages} {174511}
  (\bibinfo {year} {2008})}\BibitemShut {NoStop}%
\bibitem [{\citenamefont {Annett}(2004)}]{Annett_Book}%
  \BibitemOpen
  \bibfield  {author} {\bibinfo {author} {\bibfnamefont {J.~F.}\ \bibnamefont
  {Annett}},\ }\href@noop {} {\emph {\bibinfo {title} {Superconductivity,
  {S}uperfluids and {C}ondensates}}}\ (\bibinfo  {publisher} {Oxford University
  Press},\ \bibinfo {address} {Oxford},\ \bibinfo {year} {2004})\BibitemShut
  {NoStop}%
\bibitem [{\citenamefont {Ketterson}\ and\ \citenamefont
  {Song}(1999)}]{KettersonSong}%
  \BibitemOpen
  \bibfield  {author} {\bibinfo {author} {\bibfnamefont {J.~B.}\ \bibnamefont
  {Ketterson}}\ and\ \bibinfo {author} {\bibfnamefont {S.~N.}\ \bibnamefont
  {Song}},\ }\href@noop {} {\emph {\bibinfo {title} {Superconductivity}}}\
  (\bibinfo  {publisher} {Cambridge University Press},\ \bibinfo {address}
  {Cambridge},\ \bibinfo {year} {1999})\BibitemShut {NoStop}%
\bibitem [{\citenamefont {Covaci}\ and\ \citenamefont
  {Marsiglio}(2006)}]{Covaci_PRB73_014503}%
  \BibitemOpen
  \bibfield  {author} {\bibinfo {author} {\bibfnamefont {L.}~\bibnamefont
  {Covaci}}\ and\ \bibinfo {author} {\bibfnamefont {F.}~\bibnamefont
  {Marsiglio}},\ }\href@noop {} {\bibfield  {journal} {\bibinfo  {journal}
  {Phys. Rev. B}\ }\textbf {\bibinfo {volume} {73}},\ \bibinfo {pages} {014503}
  (\bibinfo {year} {2006})}\BibitemShut {NoStop}%
\bibitem [{\citenamefont {Balian}\ and\ \citenamefont
  {Werthamer}(1963)}]{Balian_PhysRev131_1553}%
  \BibitemOpen
  \bibfield  {author} {\bibinfo {author} {\bibfnamefont {R.}~\bibnamefont
  {Balian}}\ and\ \bibinfo {author} {\bibfnamefont {N.~R.}\ \bibnamefont
  {Werthamer}},\ }\href@noop {} {\bibfield  {journal} {\bibinfo  {journal}
  {Phys. Rev.}\ }\textbf {\bibinfo {volume} {131}},\ \bibinfo {pages} {1553}
  (\bibinfo {year} {1963})}\BibitemShut {NoStop}%
\bibitem [{\citenamefont {Sigrist}\ and\ \citenamefont
  {Ueda}(1991)}]{Sigrist_RMP63_239}%
  \BibitemOpen
  \bibfield  {author} {\bibinfo {author} {\bibfnamefont {M.}~\bibnamefont
  {Sigrist}}\ and\ \bibinfo {author} {\bibfnamefont {K.}~\bibnamefont {Ueda}},\
  }\href@noop {} {\bibfield  {journal} {\bibinfo  {journal} {Rev. Mod. Phys.}\
  }\textbf {\bibinfo {volume} {63}},\ \bibinfo {pages} {239} (\bibinfo {year}
  {1991})}\BibitemShut {NoStop}%
\end{thebibliography}

\end{document}